\documentclass[a4paper,11pt, twocolumn]{article}

\usepackage{graphicx}
\usepackage{amsfonts}
\usepackage{color}
\usepackage{authblk}
\usepackage[utf8x]{inputenc}
\usepackage{amsmath}
\usepackage{placeins}
\usepackage[twocolumn,textwidth=18.31cm,columnsep=.81cm]{geometry}
\usepackage[super]{natbib}
\usepackage{abstract}

\def\<{\langle}
\def\>{\rangle}
\def\tw{t_w}
\def\twf{t_p}

\newcommand{\textgr}[1]{\textcolor{black}{#1}}

\title{Cage--jump motion reveals universal dynamics and non--universal structural features in glass forming liquids }

\author[1,2,*]{R Pastore}   
\author[1]{A Coniglio}
\author[3,1]{A de Candia}
\author[1]{A Fierro}
\author[4,1]{M Pica Ciamarra}
\affil[1]{
CNR--SPIN, sezione di Napoli,
Dipartimento di Fisica, Campus universitario di Monte S. Angelo, Via Cintia, 80126 Napoli, Italy
}

\affil[2] {
UC Simulation Center, University of Cincinnati, and Procter \& Gamble Co., Cincinnati, Ohio 45219, USA
}
\affil[3]{
Dipartimento di Fisica, Universit\'a di Napoli Federico II, Campus universitario di Monte S. Angelo, Via Cintia, 80126 Napoli, Italy
}
\affil[4]{
Division of Physics and Applied Physics, School of Physical and Mathematical Sciences, \newline Nanyang Technological University, Singapore
}

\affil[*] {Corresponding author: pastore@na.infn.it}

\date{}
\begin{document}
\twocolumn[
\maketitle
\begin{onecolabstract}
The sluggish and heterogeneous dynamics of glass forming liquids is 
frequently associated to the transient coexistence of two phases of 
particles, respectively with an high and low mobility. In the absence of 
a dynamical order parameter that acquires a transient bimodal shape, 
these phases are commonly identified empirically, which makes difficult 
investigating their relation with the structural properties of the 
system. 
\textgr{Here we show that the distribution of single particle 
diffusivities can be accessed within a Continuous Time Random Walk 
description of the intermittent motion, and that this distribution 
acquires a transient bimodal shape in the deeply 
supercooled regime, thus allowing for a clear identification of the 
two coexisting phase.} In a simple two-dimensional glass forming model, 
the dynamic phase coexistence is accompanied by a striking structural 
counterpart: the distribution of the crystalline-like order parameter 
becomes also bimodal on cooling, with increasing overlap between ordered 
and immobile particles. 
\textgr{This simple structural signature is absent in other models, 
such as the three-dimesional Kob-Andersen Lennard-Jones mixture, where 
more sophisticated order parameter might be relevant.
In this perspective}, the identification of the two dynamical 
coexisting phases opens the way to deeper investigations of 
structure-dynamics correlations.


\end{onecolabstract}
]

\clearpage
\section{Introduction}
Probably the main paradox of structural glasses is the
apparent mismatch between dynamics and structure:
the dynamics dramatically slows down on cooling, as highlighted by the huge grow of the relaxation time.
By contrast, the structure remains seemingly unchanged, insomuch as 
it is not possible to distinguish a glass from a simple liquids simply focussing
on pair correlation functions, such as the structure factor or the radial distribution function\cite{Debenedetti, RevBerthier, RevCavagna}.
However, recent results suggested that structural fingerprints of glasses can be found
in higher order geometrical motifs, known as locally preferred structures (LPS),
or more sophisticated order parameter, such as the local configurational entropy\cite{RevRoyall},
or introducing an external perturbation, such as particle pinning\cite{Verrocchio, Tarzia} or static shear\cite{Delgado}.

A key to investigate the interplay between structure and dynamics arises from the observation
of the temporary clustering of particles with similar mobility, known
as Dynamic Heterogeneities (DHs)~\cite{DHbook}.
This inspired a picture of glassy dynamics as the transient coexistence of a "slow" and of a "fast" phase,
that only mix on the timescale of the relaxation time~\cite{Schmidt1991,Glotzer,Ediger}.
Relevant insights in this direction arise from systems driven out of equilibrium
by external fields coupled to the particle mobility\cite{Chandler_Science, Cammarota2010},
where a true dynamical phase transition and a bimodaly distributed
order parameter have been found.
Interestingly, a similar result has been obtained when a chemical potential is
coupled to certain LPS, suggesting that the dynamical phase transition
may have a structural origin\cite{Speck2012}.
Conversely, in equilibrium systems the definition of dynamic phase coexistence is much more problematic,
being not supported by the identification
of a dynamic order parameter that takes two different values in the two phases;
for instance, the van-Hove distribution, often used to identify DHs has long tails, not a bimodal shape (with distinct maxima),
so that the two phases are usually empirically defined 
by fixing an arbitrary threshold on the particle displacements \cite{WeeksScience}.
In addition, correlations of local order and particle mobility have been proved to be highly system dependent
and the existence of a causal link between structure and dynamics is still far to be proved,
at least to a general extent\cite{Harrowell2004, Harrowell2006, Hocky2014}.  

Investigation of the single particle motion is promising to shed new light
on these issues, because of its well known
intermittent character. Indeed, particles in a glass formers spend
most of their time confined within the
cages formed by their neighbors, seldom making a jump to different cages
\cite{Inter1,Inter2, Makse2009, Vollmayr1, Baschnagel1, Baschnagel2, SM_review}.
From the one hand, the ability of a particle to perform a jump
is expected to be directly affected by the local structure.
From the other hand, these single particle jumps
can be thought as the building block of the macroscopic relaxation.
We have recently investigated the cage-jump motion in molecular dynamic
simulations\cite{SM14, SciRep, SM15_corr} and experiments on colloidal glasses\cite{SM15} using a parameter-free algorithm to
segment the trajectory of each particle in cages and jumps. 
We found that jumps identified in this way are short-lasting irreversible events,
thus suggesting to describe the glassy dynamics in terms of these elementary relaxation events. 
In this paper, we demonstrate that this leads to a very general and robust description of the macroscopic dynamics
of glass forming liquids and to properly define the transient phase coexistence. 

Here we show, via molecular dynamics simulations of a 2D binary mixture of soft disks, 
that the dynamics is well described within a continuous time random walk approach (CTRW) \cite{Montroll}, 
where each step coincides with a jump.
In this framework, the diffusivity of each particle at a given time is proportional to the number
of jumps it has done. 
This allows to investigate the distribution of diffusivities focusing on
the distribution of the number of jumps per particle. In the deeply
supercooled regime, this distribution temporarily acquires a bimodal shape, 
and thus allows for the clear identification of two phases of mobile and immobile particles,
confirming previous results for a prototypic 3D model of glass forming liquids \cite{SciRep}.  
In the present 2D model, the dynamic phase coexistence is accompanied by a striking structural counterpart: 
 the distribution of a simple crystalline-like order parameter is also bimodal, the more the lower the temperature, 
 and the overlap between the most ordered and the slowest particles increases on cooling.
Such a simple and clear structural signature is not a universal feature of glass formers,
and more sophisticated structures, if any, might be relevant in other cases.
In  this respect, the possibility to identify two dynamical particle populations, without introducing any arbitrary threshold, 
has the general consequence to facilitate further investigations of structure-dynamics relationships.

\section{Methods}
\label{sec:methods}
We have performed NVT molecular dynamics 
simulations~\cite{LAMMPS} of a 
50:50 binary mixture of $N = 10^3$ of particles
interacting via an Harmonic potential, $V(r_{ij}) = \epsilon \left((\sigma_{ij}-r_{ij}) 
/\sigma_L\right)^2 \Theta(\sigma_{ij}-r_{ij})$, in two dimensions.
Here $r_{ij}$ is the interparticle separation and $\sigma_{ij}$ the average diameter of the interacting particles.
We consider a mixture of particles with a diameter ratio 
$\sigma_{L}/\sigma_{S} =1.4$,
known to inhibit crystallization, at a fixed area fraction $\phi = 1$. 
Units are reduced so that $\sigma_{L}=m=\epsilon=k_B=1$, 
where $m$ is the mass of both particle species and $k_B$ the Boltzmann's 
constant. The two species behave in a qualitatively analogous way,
and all data we have presented refer to the smallest component, 
but for those of Fig.~\ref{fig:panel2} that consider both species.
The trajectory of each particle is segmented in a series of cages
interrupted by jumps using the algorithm of Ref.~\cite{SM14}:
we associate to each particle, at each time $t$,
the fluctuations $S^2(t)$ of its position computed
over the interval $[t-10t_b:t+10t_b]$, with
$t_b$ ballistic time. At time $t$, a particle is considered
in a cage if $S^2(t) < \<u^2\>$, as jumping otherwise.
Here $\<u^2\>$ is the temperature dependent Debye--Waller factor,
we determine from the mean square displacement as in
Ref.~\cite{Leporini}.
We note that in this approach jumps are finite lasting events, as
the time at which each jump (or cage) starts and ends is identified  
by monitoring when $S^2$ equals $\<u^2\>$.
Accordingly, we have access to the time, $\twf$, a particle
persists in its cage before making the first after an an arbitrary time origin $t=0$.  (persistence time),
to the waiting time between subsequent jumps of the same particle $\tw$ (cage duration), 
and to the duration $\Delta t_J$ and the length $\Delta r_J$ of each jump.

\section{Results}

\subsection{Dynamics}
Investigating the same model considered here~\cite{SM14},
as well as the 3D Kob-Andersen Lennard-Jones (3D KA-LJ) binary mixture~\cite{SciRep} and 
experimental colloidal glass~\cite{SM15}
we have previously shown that the jumps identified using the algorithm described in Sec.\ref{sec:methods}
are truly elementary relaxation events, at least in the investigated regimes.
On the one side, jumps have a small average duration $\<\Delta t_J\>$, essentially constant on cooling, 
despite the relaxation time increasing by order of magnitudes.
On the other side, jumps are irreversible events, as 
the mean square displacement increases linearly with the number of jumps per particle $n_J$,
$\<r^2(n_J)\> \propto n_J$. 
These outstanding jumps properties allow to describe the glassy dynamics 
of atomistic systems as a continuous time random walk (CTRW),
extending an approach successfully used for idealized lattice models\cite{berthier_epl2005, Chandler_SE}.
In the following of this section, first we illustrate the basic cage-jump properties needed to a CTRW description,
then we show that this approach quantitatively describes the macroscopic dynamics and allows
to access the time dependent distribution of single particle diffusivities.

\subsubsection{From cage-jumps to CTRW} 
The CTRW model and its wide range of applications have been largely described in literature~\cite{FellerBook}.
Briefly,  particle perform a stationary and  isotropic random walk,
but the lag time and the the step size fluctuate according to given distributions.
In our description, each step corresponds to a jump and,
accordingly, the temporal features of this process are fixed by the distribution $P(\tw)$ 
of the waiting time between jumps, or equivalently, 
by the distribution $F(\twf)$ of the persistence time $\twf$.
Indeed, in the CTRW framework these distributions are related through the relation~\cite{Feller,Lax}:
\begin{equation}
\label{eq:feller}
 F(\twf) = \<\tw\>^{-1} \int_{\twf}^\infty P(\tw) d\tw.
\end{equation}
This equation also predicts that the averages, $\<\tw\>$ and $\<\twf\>$,
coincide, $\<\tw\>=\<\twf\>$, if $P(\tw)$ is exponential.

\begin{figure}[t!]
\begin{center}
\includegraphics*[scale=0.36]{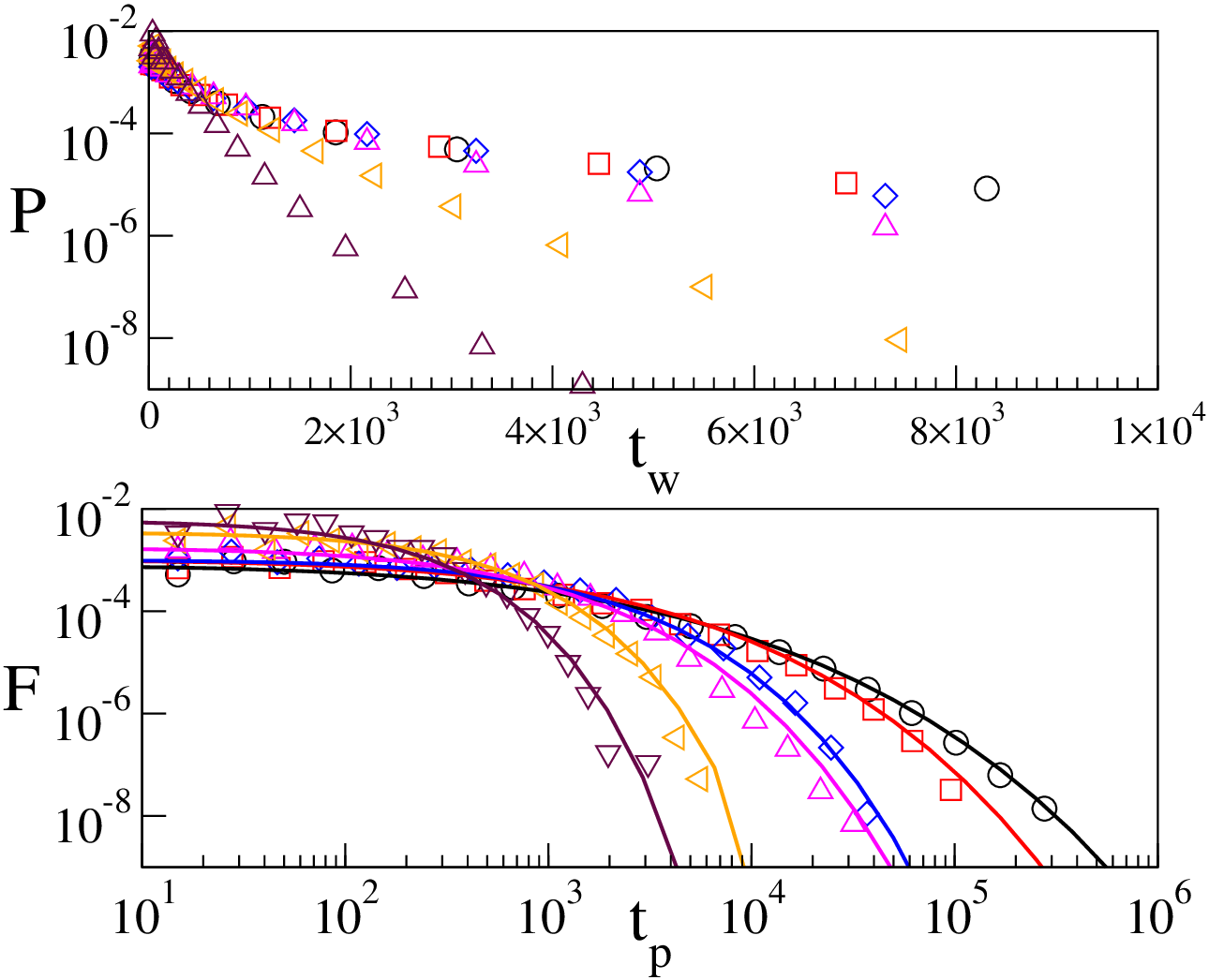}
\end{center}
\caption{\label{fig:feller}
Top panel: probability distribution of the waiting time 
between subsequent jumps of the same particle, $P(\tw)$. 
The lin-log plot clarifies that the distribution is exponential at high temperature 
and progressively deviates from this behaviour on cooling.
Bottom panel: probability distribution of the time particles persist in 
their cages before making their first jump, $F(\twf)$ (points), and prediction
obtained from the CTRW model (lines), as in Eq.~\ref{eq:feller}. 
The prediction has been obtained numerically integrating the data for $P(\tw)$
in the top panel. In both panels the temperature is $T = 2.5$, $2.2$, $2.0$, $1.9$, $1.8$, and 
$1.7 \times 10^{-3}$, from left to right.
}
\end{figure}

Figure \ref{fig:feller} shows $P(\tw)$ and $F(\twf)$ for different temperatures. 
The top panel clarifies that, at high temperature, the decay of $P(\tw)$ is compatible with an exponential,
 as it occurs when jumps originate from a Poissonian process.
At low temperature, instead, the distribution becomes clearly non-exponential,
evidencing the growth of temporal correlations.
As reported in Ref.\cite{SM14, Leporini_PRE}, other functions, 
such as power laws with exponential cutoff or stretched exponentials,
provide reliable fits to $P(\tw)$ over the whole range of investigated temperature. 
The bottom panel compares $F(\twf)$  with the prediction from Eq.\ref{eq:feller}, 
demonstrating a very good agreement.

\begin{figure}[t!]
\begin{center}
\includegraphics*[scale=0.36]{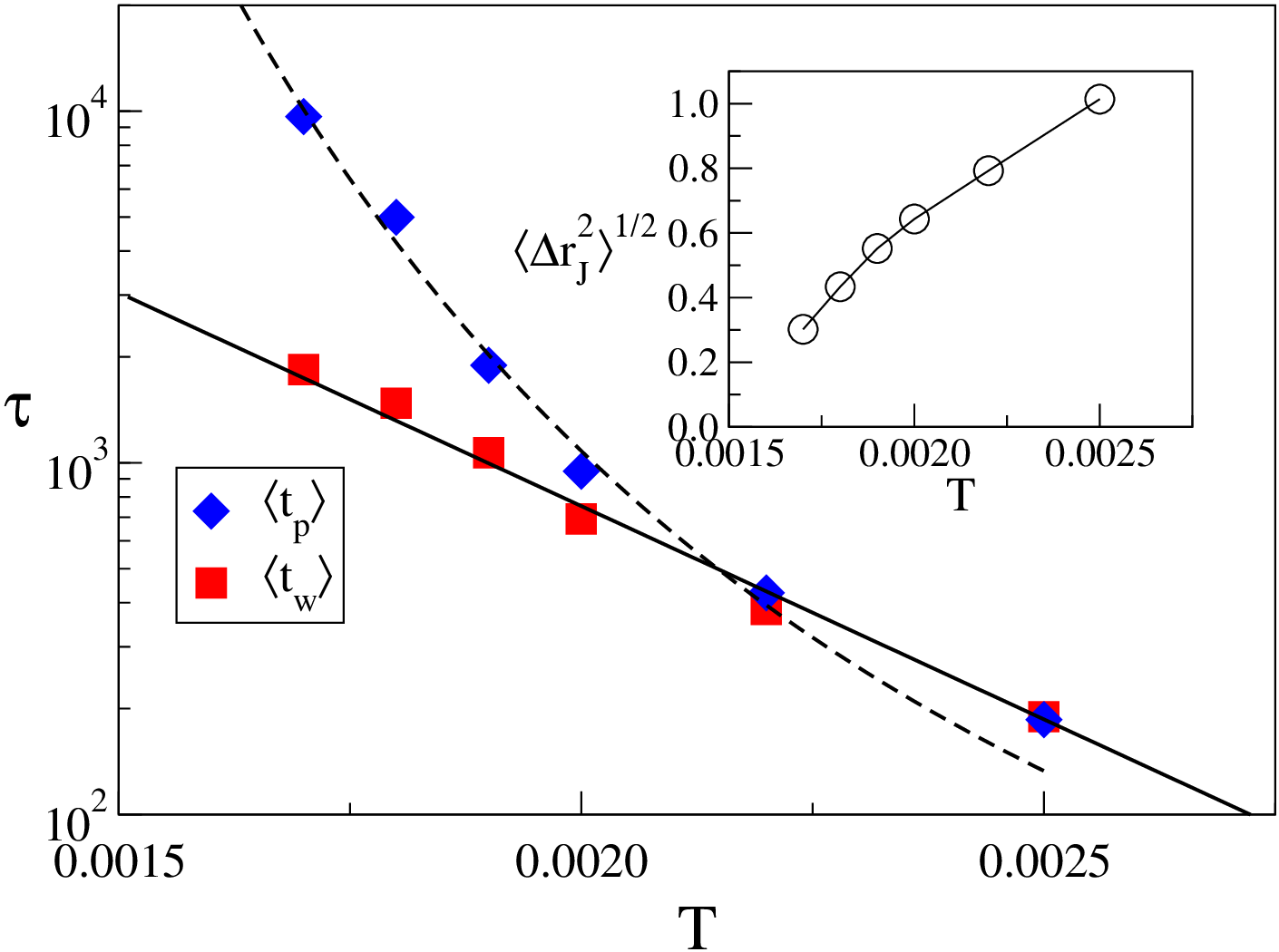}
\end{center}
\caption{\label{fig:times}
Temperature dependence of the average time particles persist in a cage before 
making the first jump, $\<\twf\>$,
and of the average cage duration, $\<\tw\>$. 
$\<\tw\>$ is well described by an Arrhenius $\<\tw\> \propto \exp\left(A/T\right)$ (full line).
$\<\twf\>$ grows \'a super--Arrhenius law. The dashed line is a fit to
$\<\twf\> \propto \exp\left(A/T^2\right)$ but other functional forms, including the Vogel--Fulcher one,
also describe the data.
The inset illustrates the temperature dependence of the average jump length. The line is a guide to the eye.
}
\end{figure}

Figure \ref{fig:times} (main panel) shows the the fundamental timescales
of the CTRW approach, $\<\tw\>$ and $\<\twf\>$, as a function of the temperature.
As predicted by Eq.~\ref{eq:feller}, $\<\tw\>=\<\twf\>$ at high temperature,
where the waiting time distribution is indeed nearly exponential.
Upon cooling, instead, the growth of temporal correlations leads to a decoupling between the two timescales.
In particular, we find that $\<\tw\>$ grows \`a la
Arrhenius, while $\<\twf(T)\>$ increases with a faster 
super--Arrhenius behavior.
This can be understood considering that the waiting times 
from particles performing subsequent rapid jumps have a major weight on $\<\tw\>$ 
but none on $\<\twf\>$.  Indeed this latter takes contributions only by 
the first jump of each particle and is therefore more affected by rare but very long waiting times.

A comparison with experimental results 
suggests that $\<t_w\>$ and $\<\twf\>$ correspond respectively to the $\beta$ and 
to the $\alpha$ relaxation time scales of structural glasses~\cite{Debenedetti, berthier_epl2005, Hedges2007},
as we better motivate later on. 
Furthermore, the temperature $T_x\simeq0.002$ where  $\<t_w\>$ and $\<\twf\>$ firstly
 decouples marks the beginning of the Stokes-Einstein (SE) breakdown
at the wavelength of the jump length\cite{SM15_corr}, and, is akin of the onset temperature \cite{Chandler_PRX}.

Finally, the spatial features of the CTRW approach are fixed by the mean square jump length, $\<\Delta r_J\>$ which
decreases on cooling as illustrated in Fig.\ref{fig:times} Inset.

\subsubsection{CTRW predictions of macroscopic dynamics}
We start showing that the macroscopic relaxation can be related to the statistical features of cage--jump motion
focusing on the persistence correlation function, we define as the fraction $p$
of particles that has not jumped 
up to time $t$~\cite{Chandler_PRE, PRL2011, Fractals, Chaudhuri}.
\begin{figure}[t!!]
\begin{center}
\includegraphics*[scale=0.36]{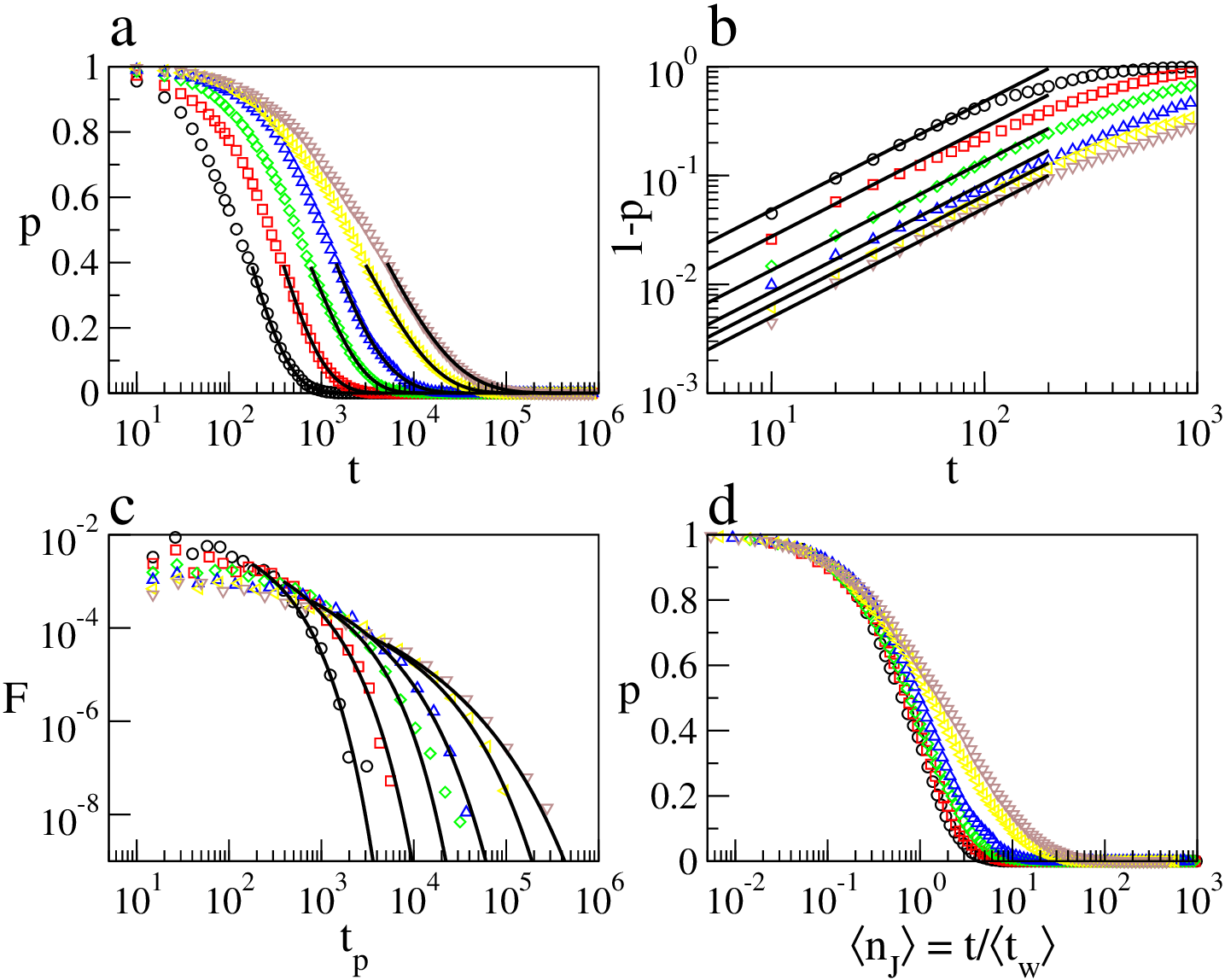}
\end{center}
\caption{\label{fig:p_njumps}
Panel a shows the decay of the persistence, together with stretched 
exponential fits of its long time behavior. 
Panel b illustrates the early time decay of the persistence. Lines
correspond to $1-p(t) = t/\<\tw\>$.
Panel c compares $F(\twf)$ with full lines derived
from the stretched exponential fit to the persistence, see text.
Panel d illustrates the dependence of the persistence on the average number of 
jumps per particle.
In all panels the temperature is $T = 2.5$, $2.2$, $2.0$, $1.9$, $1.8$, and 
$1.7 \times 10^{-3}$, from left to right.
}
\end{figure}
According to the CTRW framework, the persistence can be related to the distribution
of the time $\twf$ particles persist in their cages before making the first jump,
$p(t)=1-\int_{\twf=0}^{t} {F(\twf) d\twf}$ ~\cite{berthier_epl2005, Hedges2007, Chandler_SE}. 
The relation between $p$ and $F$ is illustrated in Fig.~\ref{fig:p_njumps}.
At short times all jumps contribute to the decay of the persistence;
consistently, we find $p(t)=1-t/\<\tw\>$, as $\<\tw\>^{-1}$
is the rate at which particles jump, and $F = -dp(t)/dt = \<\tw\>^{-1}$. 
At long times we observe the persistence to decay with a 
stretched exponential, $p(t) \propto \exp\left(-(t/\tau)^\beta\right)$,
and therefore expect $F(\twf) = -dp(t)/dt \propto \tau^{-\beta} t^{\beta-1} \exp \left(-(t/\tau)^\beta 
\right)$, as verified in Fig.~\ref{fig:p_njumps}c.
As usual in structural glasses, the stretched
exponential exponent $\beta(T)$ decreases on cooling
from its high temperature value, $\beta = 1$.
This decrement causes changes in the macroscopic dynamics,
which can be highlighted considering the decay of the persistence as 
a function of the average number of jumps per particle,
 $\<n_J(t)\> = t/\<\tw\>$, as in Fig.~\ref{fig:p_njumps}d. 
The different curves collapse at small $\<n_J\>$,
as the short time dynamics is controlled by $\<\tw\>$.
Conversely, at a given average large number of jumps per particle,
the lower the temperature the higher the value of the persistence,
indicating the presence of growing temporal correlations between the jumps,
with particles that have already jumped being more likely 
to perform subsequent jumps. 
In this respect, it is worth noticing that the CTRW approach neglects spatial correlations but encodes the presence of
temporal heterogeneities in the form of the waiting time distributions.

\begin{figure}[t!]
\begin{center}
\includegraphics*[scale=0.36]{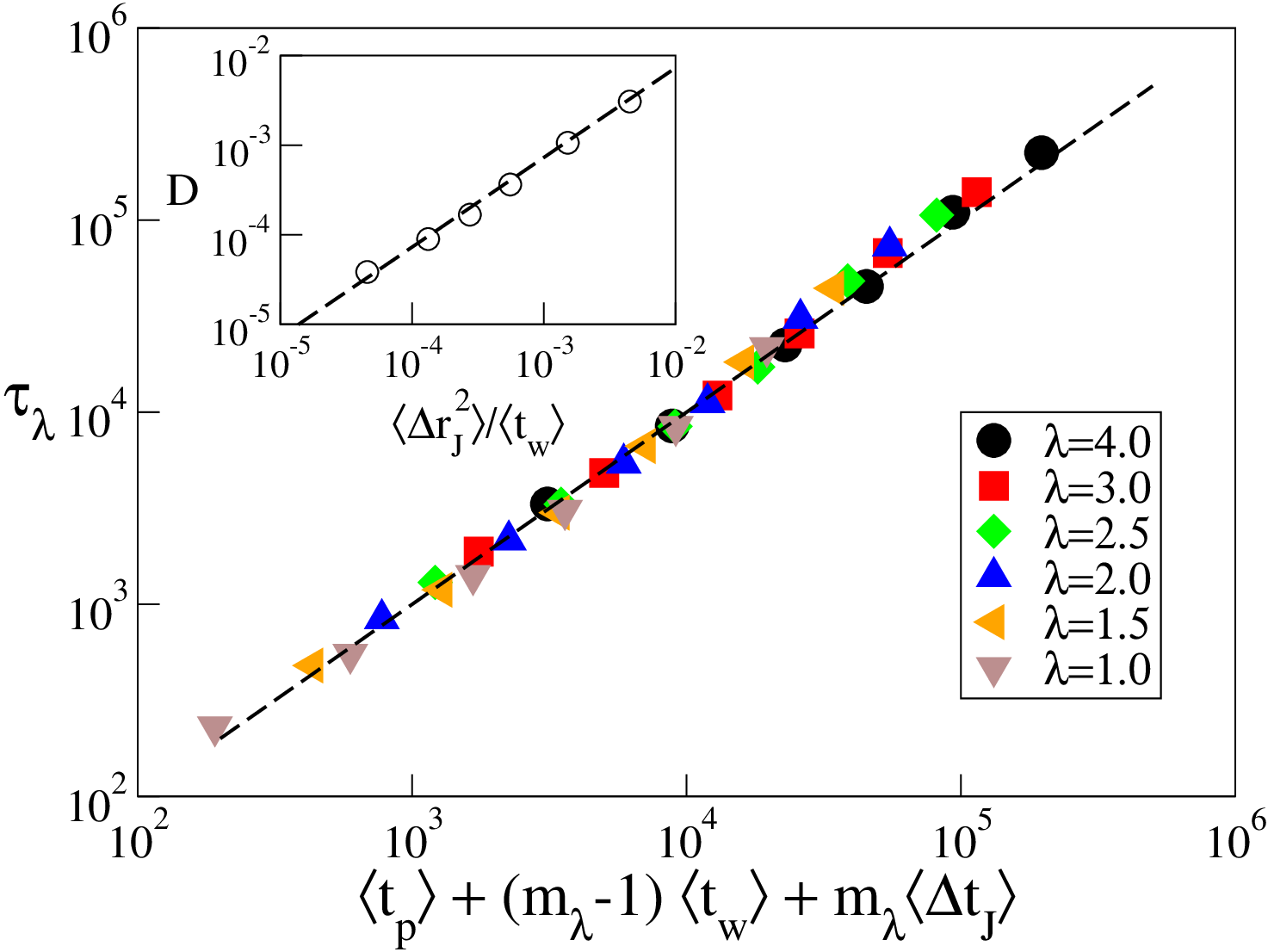}
\end{center}
\caption{\label{fig:tau_a}
The relaxation time $\tau_\lambda$ at length scale $\lambda$
is directly related to the cage--jump properties, as in Eq.~\ref{eq:t_l}.
$\lambda$ is measured in unit of the diameter of the largest particles.
For each value of $\lambda$ we compare values of $\tau_\lambda$
measured at six different values of the temperature.
The inset illustrates the diffusivity versus its 
prediction in terms of the cage jump properties.
}
\end{figure}

The diffusion coefficient and the relaxation time of the self intermediate scattering functions (ISSF),
 which are commonly measured in experiments,
can be also related to the statistical features of the cage--jump motion.
For the diffusion constant $D$, the CTRW approach predicts 
$D = \< \Delta r_J^2 \> /\<t_w\>$, as illustrated in Fig.~\ref{fig:tau_a}(inset). 
The ISSF relaxation time $\tau_\lambda$ 
at a generic wavelength $\lambda$ (wavevector $2\pi/\lambda$),
is expected to equals the average time a particle
needs to move a distance $\lambda$, as 
in the CTRW approach this is simply the time particles need to perform, on average, $m_\lambda(T) = 
\lambda^2/\<\Delta r^2_J(T)\>$ jumps. 
Accordingly, this time is fixed by the average first jump waiting time, $\< \twf \>$,
by the average cage duration, $\< \tw \>$, and, by the average 
jump duration, $\<\Delta t_J\>$,
\begin{equation}
\label{eq:t_l}
\tau_\lambda \propto \<\twf\> + (m_\lambda-1)\<\tw\> + m_\lambda\<\Delta t_J\>.
\end{equation}
The last term is actually negligible at low temperatures, where $\<t_w\> \gg 
\<\Delta t_J\>$~\cite{SM14}.
Fig.~\ref{fig:tau_a} shows that this prediction
agrees very well with the 
measured data in the investigated range of $\lambda$, with a coefficient
of proportionality of the order of $1$. We remark that we have explored
values of $\lambda$ ranging from $1$ to $4$ diameters of the largest component, 
corresponding to a relaxation time varying more than three decades.

Exploiting the dependence of the diffusivity and of the relaxation time
on the jump properties,
the breakdown
of the Stokes--Einstein relation $\tau_{\lambda} \propto D^{-1}$ appears to be mainly
controlled by the ratio of $\< \twf(T) \>$ and $\< \tw(T) \>$, 
but it is also affected by the temperature dependent jump length,
differently to lattice models, where the step size is fixed.
As a consequence, by equating the first two terms of the r.h.s. of Eq.~\ref{eq:t_l},
the length scale below which the breakdown of the SE relation occurs ~\cite{berthier_epl2005},
$l \simeq \<\Delta r^2_J\>^{1/2} (\< \twf \>/\< \tw \>)^{1/2},$
also depends on the jump length.

The relation of $\< \twf(T) \>$ and $\< \tw(T) \>$ with the macroscopic dynamics,
supports their correspondence with the $\beta$ ($\<t_w\>$) and 
the $\alpha$ ($\<\twf\>$) relaxation timescales of structural glasses, as mentioned above. \\

\subsubsection{Distribution of single-particle diffusivities}
It is not possible to identify the dynamical phases 
from the van-Hove (vH) distribution of particle displacements 
(i.e. the probability that a particle has moved of a distance $r$
along a fixed direction) without introducing some empirical
criteria, for instance by  considering as `fast' $5\%$ particles with the largest displacement\cite{WeeksScience}.
This is because, if a fraction of the particles
has a given diffusivity, then their vH distribution is
a Gaussian distribution centered in zero, 
with variance proportional to the diffusivity.
In the presence of two groups of particles with different diffusivities,
the overall vH distribution is the weighted sum of two Gaussians with zero average,
and it displays a single one maximum in zero, not the two ones required to a clear identification of the two phases.  
This clarifies that, in order to investigate whereas two or more dynamical phases coexist, one should
investigate the diffusivity distribution, not the vH distribution.
However, direct measurements of the diffusivity distribution are difficult.
Furthermore, a direct inversion of the vH distribution can only be adopted\cite{Granick, Sengupta},
if the diffusivity distribution is assumed to be time independent, as in systems with a unique and sharply defined timescale.
Unfortunately this is not the case of glasses and different approaches are needed to tackle this problem.

\begin{figure}[t!]
\begin{center}
\includegraphics*[scale=0.36]{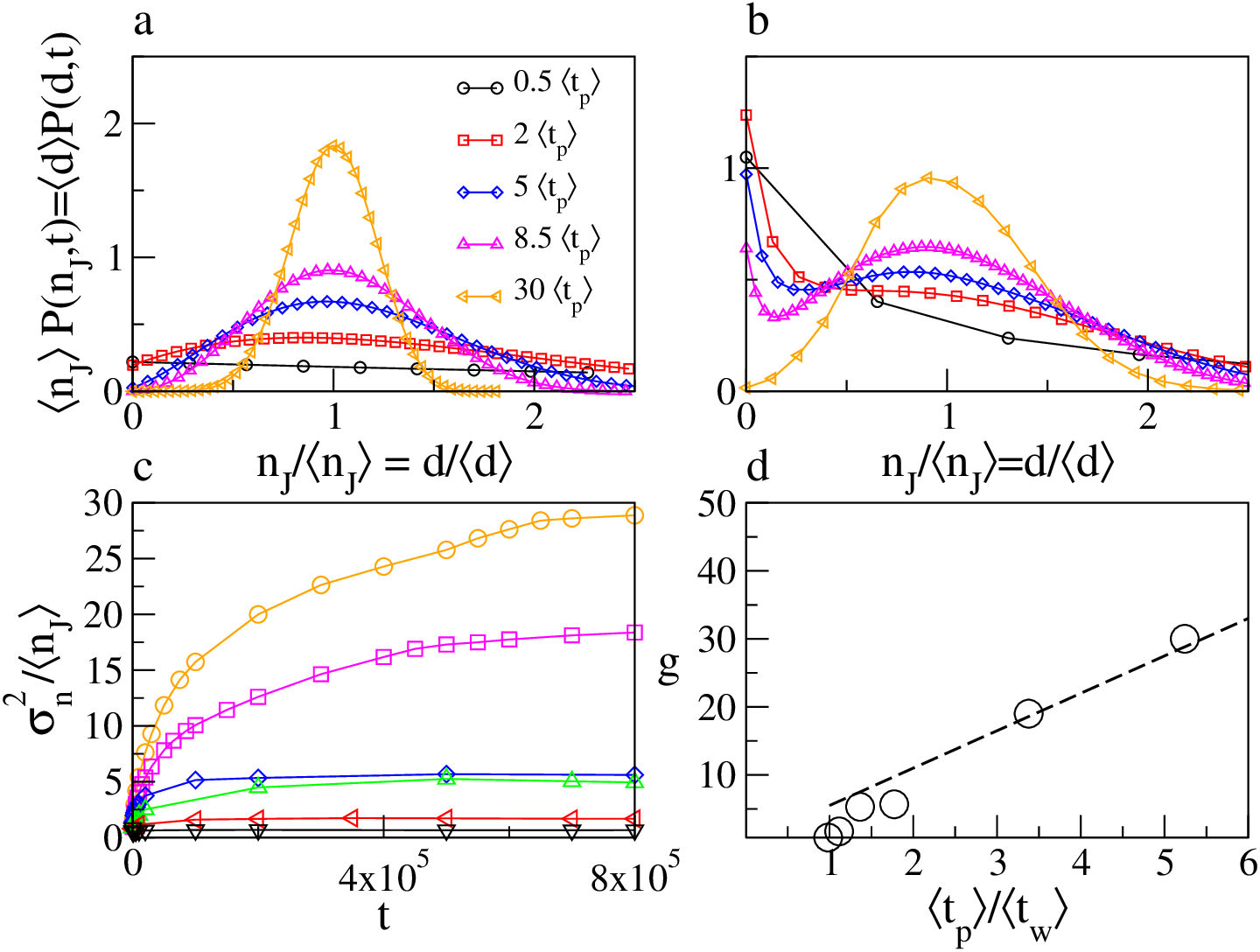}
\end{center}
\caption{\label{fig:panel1}
Panels a and b show the probability distribution
of the number of jumps per particle rescaled by the average number of jumps,
at $T=2.5\times10^{-3}$ and at $T=1.7\times10^{-3}$, respectively, and different times, as indicated.
This distribution equals that of the single particle diffusion coefficient at different
time, rescaled by the average diffusivity, as reported on the axes.
Panel c illustrates the time evolution of the variance to mean ratio of this distribution. 
Different curves refer to different temperatures, as in Fig.~\ref{fig:p_njumps}, from bottom to top. 
Panel d is a plot of the asymptotic value of the variance to mean ratio, $g$, as a function of $\<\twf\>/\<\tw\>$.
The dashed line is a guide to the eyes, $g\propto\<\twf\>/\<\tw\>$
}
\end{figure}

In the CTRW approximation, the diffusivity of particles that have performed $n_J$ jumps
at time $t$ is $d(n_J,t) = n_J(t) \<\Delta r^2_J\>/t$.
The diffusivity is therefore simply proportional to the number of jumps per unit time,
which can be easily monitored using our algorithm.
Indeed, Fig.s~\ref{fig:panel1} illustrates several features
of the distribution of the number of jumps per particle rescaled by the average number of jumps
$\<n_J(t)\> = t/\<\tw\>$, which coincides with the distribution
of the single particle diffusion coefficient
normalized by the average diffusion coefficient, $\<d\> = D$.
In Fig.s~\ref{fig:panel1}a and b we show these distributions
at different times, respectively at the highest and lowest temperature we have considered.

At $t = 0$, $P(n_J;0) = \delta(n_J)$  ($P(d;0) = \delta(d)$),
as particles have not jumped; conversely, in the infinite
time limit the distributions have a Gaussian shape with average value
$\<n_J(t)\>$ ($d = \<d\>$).
At high temperature, the distribution gradually broadens in time,
and its maximum moves from $n_J \simeq 0$ to $n_J \simeq \<n_J\>$.
At low temperature, conversely,
the distribution becomes bimodal, before reaching its asymptotic shape. 
This signals the coexistence of
an immobile or low-diffusivity phase with particles having performed few or no jumps at all,
and of a mobile or high-diffusivity phase, with particles having performed many jumps.
We rationalize the appearance of a bimodal shaped distribution considering that 
$P(n_J;t)$ is affected by the two timescales, $\<\twf\>$ and $\<\tw\>$.
The slow timescale, $\<\twf\>$, 
controls the value of the peak at $n_J = 0$, that equals the persistence correlation function, $P(n_J=0;t) = p(t)$.
The fast timescale, $\<\tw\>$, controls the average value of the
distribution, as the position of the second maximum asymptotically occurs at 
$n_J = t/\<\tw\>$. 


Figure~\ref{fig:panel1}c shows that the variance to mean ratio $\sigma^2_n/\<n_J\>$ of $P(n_J;t)$
approaches a plateau value at long time. We found that this time
scales as $\<\twf\>^a$, with $a\simeq1.3$ (in general, $a\geq1$ and it is model dependent)\cite{SciRep}.
The plateau value $g$ also  grows on cooling.
This is an other consequence of the growing temporal correlations controlled by
the decoupling between $\<\twf\>$ and $\<\tw\>$. 
Indeed,  the CTRW approach predicts that asimptotically
$g \propto \<\twf\>/\<\tw\>$, consistently with our data (see Fig.~\ref{fig:panel1}d) .

It is worth noticing that a distribution with the long--time features of $P(n_J;t)$, i.e. a Gaussian 
distribution with variance $\sigma^2_n = \<n_J\>g$,
is obtained by randomly assigning the jumps to the particles, in group 
of $g$ elements.
Consistently, at the highest temperature, where
correlations are negligible, $g = 1$ and $P(n_J;t)$ corresponds to that 
obtained by randomly assigning each jump to the particles,
i.e. a Poisson distribution. 
The increase of $g$ on cooling indicates
that at low temperature one might observe, in the same time interval, some particles
to perform $g$ jumps, and other particles to perform no jumps at all,
which illuminates the relation $g \propto \<\twf\>/\<\tw\>$.
We remark that $g$ is  a stationary quantity,
as it holds constant in the infinite time limit. 

\subsection{Structure-dynamics correlations}
We now show that,  in the investigated system, the dynamical phase coexistence
is accompanied by clear structural changes on cooling.
These structural changes has been investigated focusing on the  hexatic order parameter
\cite{Tanaka}, as the hexatic order is that expected to be relevant
in our two dimensional system.
The hexatic order parameter of particle $j$, $\psi_6^j$, is defined as:
\begin{equation}
\label{psi}
\psi_6^j=\frac{1}{z_j} \sum_{m=0}^{z_j} e^{i6\theta^{j}_{m}},
\end{equation}
where the sum runs over the $z_j$ neighbours of the particle $j$.
We define two particles $m$ and $j$ to be neighbours if in contact,
$|r_{mj}|< \sigma_{mj}$, where $\sigma_{mj}$ is their average diameter.
$\theta_m$ is the angle between ${\bf r}_{mj}$ and a fixed direction (e.g. the $x$-axis).
$|\psi_6^j(t)|$ has its maximum,  $|\psi_6^j|=1$, if  the 
neighbours of the particle $j$ are in a hexagonal order at time $t$,
while $|\psi_6^j|=0$ for a random arrangement.
We also associate to each particle its 
averaged order parameter, $\< | \psi_6^j | \>_t$, where the average is computed
in the time interval $[0:t]$.
Note that we have considered the hexatic order parameter
to identify structural heterogeneities, as this is 
the ordered characterizing two dimensional assemblies of
soft disks. In different dimensionalities and for different potentials
other order parameters, or the excess entropy, might be more appropriate~\cite{Tanaka}.

\begin{figure}[t!]
\begin{center}
\includegraphics*[scale=0.36]{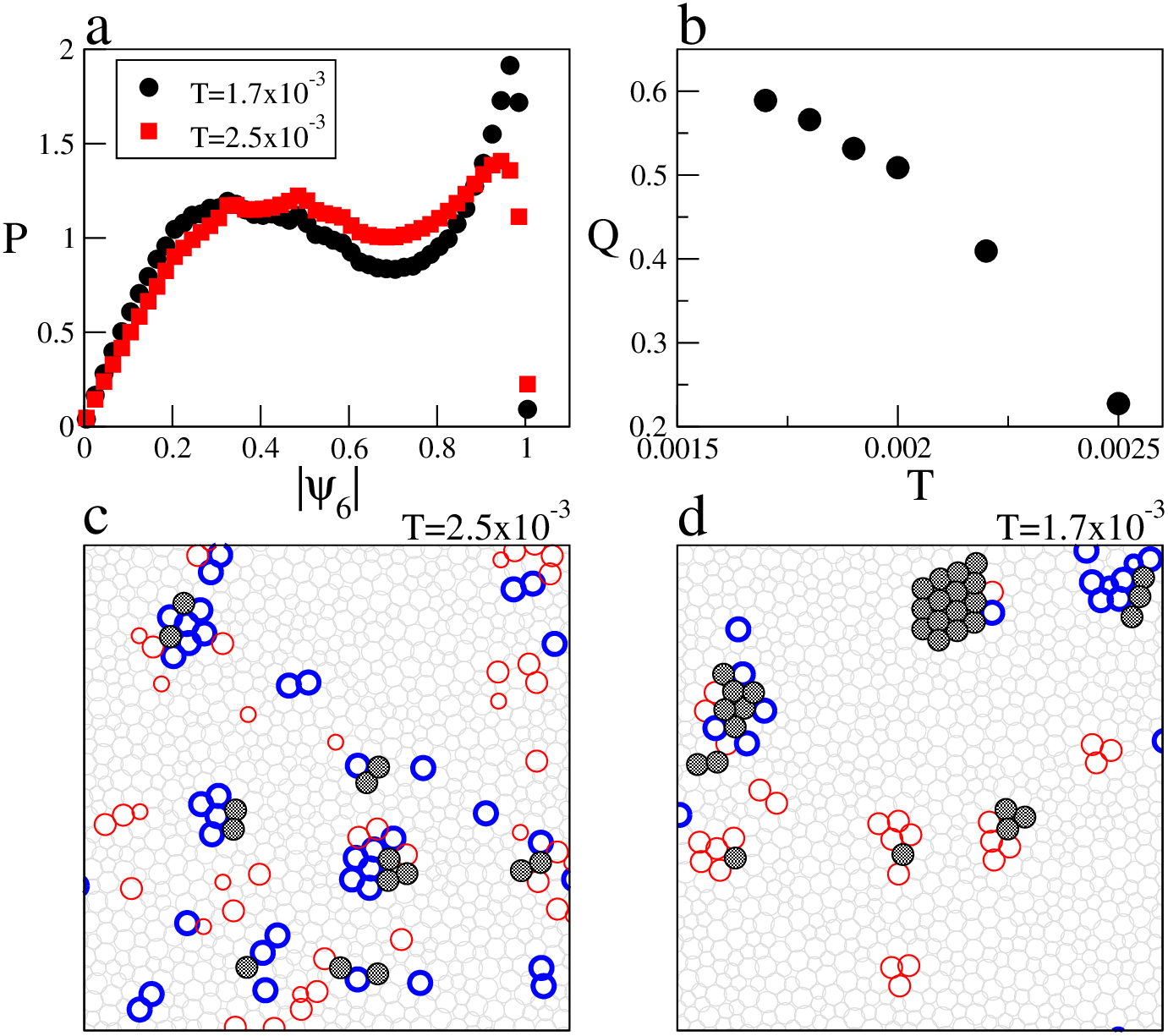}
\end{center}
\caption{\label{fig:panel2}
(a) Probability distribution of the hexatic order parameter, $P(|\psi_6|)$.
(b) Temperature dependence of the average overlap between persistent and ordered 
particles.
(c,d) Snapshots of a portion of the system at different temperatures, as indicated,
and at times $t$ such that $p(t) = 5\%$.
The $5\%$ particles that have not yet jumped (thin red circles),
and the $5\%$ with the highest averaged (from $0$ to $t$) hexatic order 
parameter (thick blue circles) are shown.
Particles being simultaneously ordered and persistent are highlighted (filled circles).
}
\end{figure}

As illustrated in Fig.~\ref{fig:panel2}a, the single--particle
hexatic order parameter has a bimodal distribution, that becomes more apparent on cooling.
This signals the presence of a small fraction of highly ordered particles,
and suggests that ordered and persistent particles might be related.

We show that this is actually the case
investigating an overlap function $Q$ between ordered and persistent particles.
In particular, we investigate the correlations between the set containing the $m=5\%$ most
persistent particles and that containing the $m=5\%$ most ordered particles.
The persistent particles set contains all the particles
that have not jumped up to a time $t$, so that $p(t) = m$.
The most ordered particles are those with the highest value
of averaged order parameter $\< |\psi_6^j| \>_t$.
For the persistent particles, $n_{p}^{i} = 1$.
Similarly, we introduce a scalar $w_{\psi}^i$ to distinguish between
the particles that are among the most ordered, $w_p^i = 1$,
and those that are not, $w_\psi^i = 0$. 
The overlap between ordered and persistent particles is therefore defined as
\begin {equation}
\label{Eq: Q}
Q(T)=\frac{ (1/N)\sum_{i=0}^{N} n_{p}^{i} w_{\psi}^i - m^2}{m-m^2},
\end{equation}

This function equals $Q = 1$ 
when the two populations coincide, and $Q = 0$ when they are uncorrelated.
Fig.~\ref{fig:panel2}b shows that $\<Q(T)\>$ increases on cooling,
and thus demonstrates the existence of strong correlations between statics and dynamics~\cite{Tanaka}.
Snapshots of the system in Fig.s~\ref{fig:panel2}c,d  
illustrate the meaning of $Q$,
confirm that ordered and persistent particles largely overlap at low 
temperature, and also show that they form clusters.
Since particles belonging to the core of these clusters can only jump
after those of the periphery, the relaxation of these clusters
is highly cooperative.
At the level of the single particle cage--jump motion, 
this k--core~\cite{kcore} like mechanism appears the one responsible
for the decoupling of the two timescales, and for the emergence of dynamical heterogeneities.
It is worth noticing that these highly ordered clusters look like micro-crystals,
which are indeed likely to form in this systems, although full scale crystallization is suppressed.

\section{Discussion}
The CTRW approach completely neglects the presence of spatial correlations between jumps of different particles.
Indeed, we recently showed that, for the model considered here, jumps actually occur in group of few particles\cite{SM15_corr}. 
Moreover, for a correct CTRW description it is crucial to fairly identify
elementary irreversible events. In the deeply supercooled regime this could become problematic,
as subsequent jumps of a same particle are expected to become increasingly anti-correlated,
in the form of back and forward moves. 
\textgr{In that case, alternative definitions of local relaxation events or more sophisticated
methods to distinguish irreversible jumps are needeed to properly describe the dynamics
within a CTRW approach.
For the  system considered here, our algorithm successfully identifies irreversible jumps, 
at least in the temperature range allowed by simulations.
Indeed, we have shown that the CTRW framework describes surprisingly well the dynamics of this atomistic
glass-formers down to the lowest investigated temperature.
In the 3D KA-LJ, small deviations from the CTRW predictions are only found at the 
lowest investigated temperature~\cite{SciRep}.}

Differences between glassy dynamics in two and three dimensions, whether fundamental or apparent,
has been debated in very recent works~\cite{Szamel15, Shiba15}, pointing out that dimensionality and system size 
lead to quantitative change in the cage-jumps properties.
Despite of these differences,  we have shown that our approach provides the same dynamical scenario in 2-D and 3-D systems, and
a robust framework to connect the microscopic and the macroscopic dynamics.

The main success of this approach is the identification of two dynamical phases trough 
the bimodal shape of the diffusivity distribution.
In the investigated 2D system, the dynamical phase coexistence 
is accompanied by a striking "structural bimodal" of the hexatic order parameter, 
that is probably related to the presence of micro-crystals.
The same clear structural signature is not observed in other standard glass formers,
where local crystallization is completely suppressed. 
This can raise the suspect that the "dynamic bimodal" distribution of diffusivities is
directly triggered by that of the hexatic order, and thus it is a peculiar property of this systems as well.
By contrast, we can claim that the temporary bimodal shape of the diffusivity distribution is a general property of glass formers,
as for the standard 3D KA-LJ models we do observe a dynamical bimodality but not a structural bimodality\cite{SciRep}, 
at least when the structure is investigated via crystalline-like order parameters. 
This demonstrates that the dynamical phase coexistence is not necessarily related to local crystallization, but does not exclude 
that correlations with different structural properties could exist.
In general, it remains an open question whether the dynamical phase coexistence has a structural counterpart,
or whether the reported structure-dynamics correlations must be considered as a by-product of particular models. \\

{\bf Acknowledgments}\\
We acknowledge financial support 
from MIUR-FIRB RBFR081IUK, 
from the SPIN SEED 2014 project {\it Charge separation and charge transport in hybrid solar cells},
and from the CNR--NTU joint laboratory {\it Amorphous materials for energy harvesting applications}. \\
\\


\end{document}